\documentclass{article}

\usepackage{amsmath} % for dealing with mathematics,
\usepackage{amsfonts}
\usepackage{bbm}
\usepackage{hyperref} %for linking the cross references
\usepackage{graphicx} %for dealing with figures.
\usepackage[usenames,dvipsnames,svgnames,table]{xcolor}
\usepackage{subfigure}
\usepackage{cite}
\newcommand{\KS}{\text{KS}}
\newcommand{\isep}{\,;\,}
\usepackage[normalem]{ulem}
\usepackage[margin=1in]{geometry}

\begin{document}

\title{Entrograms and coarse graining of dynamics on complex networks}
\author{%
  {\sc Mauro Faccin}$^*$,\\
  ICTEAM, Universit\'e catholique de Louvain, Belgium\\
  {$^*${Corresponding author: \href{mailto:mauro.faccin@uclouvain.be}{mauro.faccin@uclouvain.be}}}\\
  {\sc Michael T. Schaub},\\
  IDSS, Massachusetts Institute of Technology, USA\\
  Department of Engineering Science, University of Oxford, UK\\
  {\sc Jean-Charles Delvenne}\\
  ICTEAM and CORE, Universit\'e catholique de Louvain, Belgium
}

\maketitle

\begin{abstract}
  Using an information theoretic point of view, we investigate how a dynamics acting on a network can be coarse grained through the use of graph partitions.
  Specifically, we are interested in how aggregating the state space of a Markov process according to a partition impacts on the thus obtained lower-dimensional dynamics.
  We highlight that for a dynamics on a particular graph there may be multiple coarse grained descriptions that capture different, incomparable features of the original process.
  For instance, a coarse graining induced by one partition may be commensurate
  with a time-scale separation in the dynamics, while another coarse graining
  may correspond to a different lower-dimensional dynamics that preserves the Markov property of the original process.
  Taking inspiration from the literature of Computational Mechanics, we find that a convenient tool to summarise and visualise such dynamical properties of a coarse grained model (partition) is the entrogram.
  The entrogram gathers certain information-theoretic measures, which quantify how information flows across time steps.
  These information theoretic quantities include the entropy rate, as well as a measure for the memory contained in the process, i.e., how well the dynamics can be approximated by a first order Markov process.
  We use the entrogram to investigate how specific macro-scale connection patterns in the state-space transition graph of the original dynamics result in desirable properties of coarse grained descriptions.
  We thereby provide a fresh perspective on the interplay between structure and dynamics in networks, and the process of partitioning a network from an information theoretic perspective.
  To illustrate our points we focus on networks that may be approximated by both a core-periphery or a clustered organization, and highlight that each of these coarse grained descriptions can capture different aspects of a Markov process acting on the network.
\end{abstract}

\section{Introduction}

A large number of studies have been concerned with analyzing dynamical processes on complex networks, ranging from epidemic spreading to information diffusion, opinion formation and synchronization~\cite{vespignani2001spreading,motter2005net-sync,fortunato2005spreading,latora2006opinionDynamics,Boccaletti2006}.
As the structure of a network constrains the dynamical processes that act on top of it, the structural configuration of a network has repercussions for the dynamics we can observe.
However, certain dynamical features may not be immediate from a consideration of the network structure alone.
For instance, as outlined by Scholtes et
al.~\cite{ScholtesWiderPfitznerEtAl2014}, a simple graph representation may
hide certain temporal features such as a path dependence within a dynamics, or other memory effects~\cite{delvenne2015diffusion,rosvall2014memory}.

A common approach to incorporate such memory effects is to lift the state space of the dynamical model into a higher dimension, thereby creating an enlarged graphical representation of the system.
Within this lifted model, the dynamics can then again be described in simpler terms, e.g., by a standard Markov process such as a random walk.
The observed dynamics, vice versa, corresponds to a projection of this higher order model into the smaller original state space, in which each state corresponds to an atomic, measured event of the original process.
Stated differently, the measured process corresponds to a coarse graining, or aggregation of the higher order state space according to a particular partition.
Indeed, we may posit that in many instances the apparent memory in a dynamics
is a by-product of how we delineate events in the original process --- for
instance, in a network of commuter flows, we may partition physical space at varying levels of granularity which will lead to different representations of the dynamics.
Finding a parsimonious Markov chain that approximates an observed process more economically than lifting the state space to higher order sequences of the observed process, has been the topic of intense scrutiny, under various names such as Hidden Markov Models, Computational Mechanics, $\epsilon$--machines, etc., see e.g.~\cite{Grassberger1986, CrutchfieldYoung1989, CrutchfieldFeldman2003,Peixoto2015}.

Conversely, given a many-state dynamical system such as a Markovian flow circulating on a large network, a central task is often to find a reduced, coarse grained model, i.e., a simplified dynamical description that is able to preserve or approximate some aspects of interest of the full-size system. 
This is the direction we follow in this paper. 
Note that we ignore the issue of measurement error here: we assume that the network observed is an accurate depiction of the system of interest.
Our primary concern is thus not to assess whether a particular structure in the topology could have arisen due to a random fluctuation in the generation of the network, but how the mesoscopic structures in our empirically observed system impact on the dynamics, and in particular if they enable us to reduce the dynamical description of the system in an information theoretic sense.
Accordingly, we will primarily consider networks which can be naturally interpreted as describing the state space of a dynamical process such as information diffusion~\cite{Salnikov2016}.
Moreover, we will assume in the following that a certain partition of interest has been identified and we know into how many groups we want to partition the network.
There are many cases in which such a putative partition is already known to the researcher, e.g., by the presence of metadata, functional characteristics of the nodes, network building constraints etc. 

Alternatively, a putative partition might be generated by a community detection method. 
As many complex systems are far from unstructured but contain interesting connectivity patterns, including a wide range of motifs, cycles, subgroups, and hierarchical arrangements, finding such mesoscopic structures has been an important topic in the field of network science over the last decade.
Of particular interest has been the identification of so-called community structure, groups of nodes that are supposed to form \emph{elementary blocks} of the network.
Consequently, a plethora of different community detection algorithms have been developed in this field~\cite{newman2015generalized,lancichinetti2009community,Fortunato2016community,newman2004finding, Pons2005}, ranging from spectral clustering~\cite{vonLuxburg2007}, to statistical inference~\cite{hastings2006community, newman2007mixture, karrer2011dcsbm}.

\paragraph{Outline and article structure}
In order to assess how partitioning (the state space of a dynamical process on) a network will influence the projected dynamics, in this paper we introduce the \emph{entrogram}, which is inspired by notions from computational mechanics~\cite{Grassberger1986, CrutchfieldYoung1989, CrutchfieldFeldman2003}.

From a practical viewpoint, the entrogram provides an information-theoretic framework to capture general, dynamically relevant information about a stochastic process. 
It aggregates several information-theoretic quantities into a single histogram, which enables us to easily assess the effect of projecting a dynamics onto a particular partition.
In particular, the entrogram measures the entropy rate of the process and its memory, both in terms of how Markovian the ensuing dynamics is, as well as its higher order memory content, which may be important in various contexts~\cite{rosvall2014memory}.
The entrogram thereby quantifies how close a partition is to inducing a lumpable Markov dynamics, denoting an aggregate, projected dynamics that is still Markov, in information-theoretic terms.

As each partition of a network induces a different coarse grained dynamics, we can use the information-theoretic lens provided by the entrogram to assess how different partitions may lead to different relevant aggregate models, which capture different aspects of the dynamics.
We emphasize that the relevance of a reduced model will be contingent on the precise question the researcher is interested in, i.e., what aspects of the dynamics the aggregated network should capture, and therefore what quantities of the entrogram are considered most important.

This tenet is in line with the recent work of Peel et.\ al~\cite{Peel2017} on finding structural partitions in networks.
While in real networks, a \emph{ground truth} partition does not exist, it can nevertheless be of interest to assess whether certain structural properties of the network are related to some features of interest such as meta-data.
Instead of meta-data, here we are interested in the various ways structural features can relate to the dynamical process acting on top of the network.
As there is no ground truth in our context as well, the question we ask is thus whether a particular partition of the network is aligned with features of interest such as a time-scale separation, or a lumpable dynamics with small memory content.
In our examples we focus especially on the scenario in which there are two competing aggregate models of dynamical interest:
a partition into dense clusters, which confine a dynamics and may thereby lead to a time-scale separation, and a core-periphery model,
where the set of core nodes promotes fast diffusion over the network.

The remainder of the article is structured as follows.
In Section~\ref{sec:entrogram}, we begin our discussion by considering static networks endowed with a stationary random process, such as a random walk dynamics, and recall some information theoretic properties of interest for such a process.
We then introduce the entrogram as a tool to compactly describe such information-theoretic quantities.
In Section~\ref{sec:results}, we discuss how coarse graining a (first order) Markov process may be interpreted as a projection of the original state space onto a lower-dimensional (aggregated) state space with fewer states.
We highlight that such an aggregation has certain information-theoretic repercussions which can be captured by the entrogram: for instance the coarse grained dynamics will in general not possess the first order Markov property.
Indeed, depending on the chosen partition, the lower-dimensional process can
have severable features.
For example, it may have a small entropy rate, or correspond to an approximately lumpable dynamics.
We show how such dynamical features can be related to different types of structural blocks in a network,  offering a complementary perspective onto the role of network structure.
In Section~\ref{sec:community_or_lumbability_cases}, we focus on core-periphery and assortative cluster structure, as commonly found in many systems.
Through a series of synthetic and real-world examples, we discuss how these different types of aggregated models emphasize different aspects of a diffusion dynamics, and how these get reflected in their entrogram.
We conclude the manuscript in Section~\ref{sec:ccl} with a discussion and highlight some possible avenues for future work.

\section{Entrograms as information-theoretic descriptions of stationary random processes}%
\label{sec:entrogram}

\subsection{Coarse grained descriptions and the interplay between structure and dynamics}
Analysing the interplay between structure and dynamics of networks has been one of the mainstay topics in network science literature~\cite{Boccaletti2006}.
Random walks in particular have been shown to be useful tools to characterize topological properties of networks~\cite{Masuda2016}.
Think, e.g., of the celebrated Pagerank~\cite{brin1998pagerank} algorithm. 
Furthermore, as clusters naturally tend to trap a random walker, slowing down its diffusion across the network, random walk based algorithms have been used to find structural patterns and clusters in networks~\cite{Delvenne20072010,Rosvall29012008,piccardi2011lumpedMC,Masuda2016,DeDomenico2017}.
Implicitly, most of these approaches rely on the existence of a time scale separation:
there is a fast mixing of the random walker inside the community, but only a slow diffusion from community to community. 

Perhaps the most useful consequence of this time-scale separation in the dynamics is the possibility of model reduction, in which a community is replaced by an aggregated node.
A random walk dynamics on the aggregated network now serves as an approximation for the random walk on the original network.
More precisely, the total probability of presence in a community in the full system, is close to the probability predicted by the Markovian random walk on the aggregated network, with a level of approximation that is uniform for all large enough times~\cite{simon1961aggregation}.
This particular notion of model reduction based on a time-scale separation was formalized by Simon and Ando~\cite{simon1961aggregation}.
It forms the basis of countless macroscopic models for natural and engineered systems, where the many microscopic variables of an exact model can be reduced to a small number of macroscopic variables, due to fast microscopic equilibration.

The presence of dense clusters within a network and the resulting time-scale separation are however far from the only scenario in which a Markov chain can be suitably aggregated through a partitioning of the nodes.
The aggregation of a Markov chain induced by an arbitrary partition of the nodes is in general not Markovian.
In fact, whether a Markov chain is \emph{lumpable}, meaning that the dynamics projected onto an aggregated network are still Markov, can be easily checked for a given putative partition~\cite{kemeny1960finite, rubino1989weak, piccardi2011lumpedMC, schaub2012markov, o2013observability} as we review below.
While exact lumpability may be relatively rare in large complex networks, an approximate lumpability of the dynamics can be of significant interest as well, i.e., partitions of the nodes that induce an approximately Markovian aggregated dynamics~\cite{piccardi2011lumpedMC,dellarossa2013profiling}.
In the following we will introduce some information theoretic tools to quantify these notions in more precise terms.

\subsection{Information theoretic descriptions of dynamics}
Let us consider a generic stationary discrete-time random process:
\begin{equation}
\mathcal Y= {[\ldots, Y_{t-2}, Y_{t-1}, Y_t, Y_{t+1}, Y_{t+2}, \ldots]}_{t \in \mathbb{Z}},
\end{equation}
where each random variable $Y_t$ takes values in a finite alphabet, i.e., can be only in one of finitely many states.
This process is fully defined by the joint probability distribution $F_{\{Y_t\}}$ over ${\{Y_t\}}_{t \in \mathbb Z}$. 
Accordingly, one may obtain the marginal distributions of arbitrary subsets, such as $\{Y_t\}$, $\{Y_t, Y_{t+1}\}$, $\{Y_t,
Y_{t+1}, Y_{t+2}\}$, compatible with $F_{\{Y_t\}}$.
By stationarity, all these distributions will be independent of $t$.

If all random variables $Y_t$ are independent and identically distributed (i.i.d.), then the process $\mathcal Y$ is called a white noise, i.i.d.\ process, or Bernoulli process~\cite{peter1982ergodicity}.
If the set of part events
$\mathcal Y_{\mathcal{P}_{-1}}= \{\ldots, Y_{t-2}, Y_{t-1}\}$
and future events
$\mathcal Y_{\mathcal{F}_{+1}} = \{Y_{t+1}, Y_{t+2}, \ldots\}$
are independent when conditioned on the present $Y_t$, the process $\mathcal Y$ is called a Markov process of first order.
The process $\mathcal Y$ is called a second-order Markov process if past
$\mathcal Y_{\mathcal{P}_{-1}}$ and future $\mathcal Y_{\mathcal F_{+2}}= \{
Y_{t+2},Y_{t+3},\ldots\}$ are independent, conditioned on two consecutive symbols $\{Y_t, Y_{t+1}\}$.
A $k$-th order Markov processes can accordingly be defined as an independence relationship conditioned on a set of $k$ consecutive random variables.

Various quantities of interest can be defined for such a general process. 
For example, the Kolmogorov-Sinai entropy, or entropy rate, is defined as the average information contained in the process per time step, 
\begin{equation}
h_{\KS}(\mathcal{Y})= 
  \lim_{k \to \infty} H(Y_{t}, \ldots, Y_{t+k-1})/k=
  H(Y_t|Y_{t-1}, Y_{t-2}, \ldots)
\end{equation}
where $H(\cdot)$ denotes the usual Shannon entropy.
As the last equality indicates, for a stationary process the entropy rate is simply equal to the entropy of every new symbol knowing the complete past.
For a proof of this fact, see Ref.~\cite{cover2012elements}.
The entropy rate offers an asymptotic characterization of the stochastic process, and is an important invariant for the classification of dynamical systems~\cite{walters2000introduction}.
It describes how much information is on average generated at each time-step (by each new symbol).
The entropy rate thereby captures how \emph{unpredictable} the process is, i.e., how much surprise there is in observing each new symbol, given the whole previous trajectory.
A Bernoulli process in which each symbol takes values in a $K$ dimensional alphabet with uniform probabilities has an entropy rate of $\log(K)$. 
A deviation from this uniform distribution, or temporal correlations between the symbols of the random process will tend to decrease the entropy rate. 
An extreme case is a deterministic process: here, the list of past symbols is sufficient to predict the next symbol with certainty, resulting in a zero entropy rate.

The entropy rate is however not enough to characterize a system completely.
In particular, the entropy rate does not allow to distinguish between a white noise process with no memory, a Markovian process with a one-step memory, and a non-Markovian process with higher order memory. 
For example, Crutchfield et al.~\cite{CrutchfieldYoung1989} consider the mutual information
$I(\mathcal Y_{\mathcal P_0}\isep\mathcal Y_{\mathcal F_{+1}})$ between the set
$\mathcal Y_{\mathcal P_0} = \{ \ldots, Y_{t-1},Y_{t}\}$ of past and present states, and the set  $\mathcal Y_{\mathcal F_{+1}}$ of future states.
They proposed the \emph{excess entropy} $I(\mathcal Y_{\mathcal P_0}\isep\mathcal Y_{\mathcal F_{+1}})$ as an
important measure for the \emph{complexity} of the process.
This measure, which has also been called stored information, effective measure complexity, or predictive information by other authors, can distinguish processes with higher order memory from Bernoulli processes or deterministic processes. 
For example it is a lower bound on the size (state entropy) of any Markovian model ($\epsilon$-machine) of the measured process. 
Following a similar reasoning we introduce in the sequel a set of information-theoretic quantities that provide a compact summary of various properties of a stochastic process.

\subsection{A compact description of stationary random processes: the entrogram}

\begin{figure}[ht!]
  \centering
  \includegraphics[width=.5\textwidth]{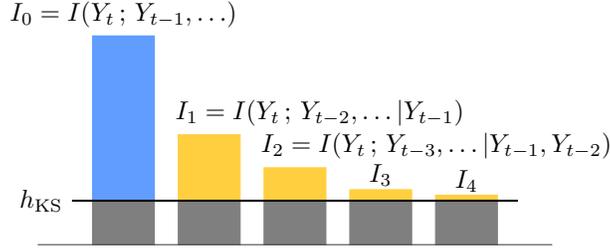}
  \caption{Entrogram.
    The plot represents the entropy fingerprint of a random process.
    The grey bottom area represents the entropy rate $h_\KS$ of the aggregated
    dynamics. The coloured upper bars represent the quantities $I_0$, $I_1$, $I_2$, etc. (see text) that describe the decay of the mutual information between temporally more and more separated symbols, when conditioning on some subsets of the past trajectory. 
    While the first blue bar represents the structural content in the
    projected dynamics, the quantities $I_1$, $I_2$, etc. (yellow), sum up to
    the higher order Markov memory $Q$, which is exactly zero for first order Markov chains. 
  }%
\label{fig:histograms}
\end{figure}

Let us consider the sequence of conditional entropies 
\begin{equation*}
\mathcal E =  \left [H(Y_t),\ H(Y_{t}|Y_{t-1}),\ H(Y_{t}|Y_{t-1}, Y_{t-2}),\ \ldots\right],
\end{equation*}
which may be suitably represented in a
histogram~\cite{CrutchfieldFeldman2003}.
We propose to call this sequence $\mathcal E$ and the corresponding histogram of entropies the \emph{extended entrogram} of the random process (see Figure~\ref{fig:histograms}).
Each entry of the extended entrogram captures how much information (uncertainty) is contained in the symbol at time $t$, when conditioning on an increasing set of previous symbols.
Note that all quantities in $\mathcal E$ are non-increasing, and converge to a final value given by the entropy rate $h_{\KS}$. 
Therefore, if we are interested only in how the information at a given time
$t$ decays towards $h_{KS}$, i.e., how the correlations in the system decay
with time, we may subtract the final value from $\mathcal E$ and obtain the
quantities $I_0= H(Y_{t})-h_{\KS}$, $I_1= H(Y_{t}|Y_{t-1})-h_{\KS}$,
$I_2=H(Y_{t}|Y_{t-1},Y_{t-2})- h_\KS$, etc.
These quantities express the information gained about the current symbol by
considering an increasingly distant past.
To see this, we can rewrite those quantities as $I_k=I(Y_t\isep Y_{t-k-1} Y_{t-k-2} \ldots |Y_{t-1}Y_{t-2} \ldots Y_{t-k})$.
The sequence of mutual information measures, 
\begin{equation*}
\mathcal{I} = [I_0, I_1,I_2,\ldots],
\end{equation*}
along with the entropy rate, is equivalent to the data of the extended entrogram above.
Therefore we will call both $\mathcal I$ and $\mathcal E$ simply the \emph{entrogram}, where the distinction to the extended entrogram should be clear from the context (see Figure~\ref{fig:histograms})

Many quantities of interest can be directly read out from the (extended) entrogram.
See Ref.~\cite{CrutchfieldFeldman2003} for an overview of more quantities that can be derived from the entrogram.
\begin{enumerate}
  \item The entropy rate $h_{\KS}$, capturing the unpredictability of the process, as discussed above.
  \item The excess entropy
      $I(\mathcal Y_{\mathcal P_0}\isep\mathcal Y_{\mathcal F_{+1}})$ discussed above turns out to be equal to the sum over the
      entrogram $I(\mathcal Y_{\mathcal P_0}\isep\mathcal Y_{\mathcal F_{+1}}) = \sum_i I_i$~\cite{CrutchfieldFeldman2003}. 
  \item The entropy $H(Y_t)$ of the stationary distribution. 
      The entropy $H(Y_t)$ measures the information content of the stationary distribution of the random process over the state space.
      For a state space with $K$ states the entropy of the stationary distribution will reach its maximum $\log(K)$ for a uniform distribution.
      The more concentrated the stationary distribution is on a certain set of states, the smaller $H(Y_t)$.
      Note that there are many processes with the same stationary distribution, and therefore the same $H(Y_t)$, which have however a very different entropy rate.
  \item The information $I_0 = H(Y_t) - h_{KS}=I(Y_t \isep Y_{t-1},Y_{t-2},\ldots)$, captures how well the current symbol can be predicted from the past trajectory (in blue on Figure~\ref{fig:histograms}).
      This quantity has been referred to as total predictability in Ref.~\cite{CrutchfieldFeldman2003}.

     Note that, if we were to encode the trajectories of the process $\mathcal Y$, the entropy rate would give a lower bound for compression achievable by an optimally efficient coding scheme, whereas $H(Y_t)$ provides a lower bound for the case where the symbols at each time-step are independent.
     $I_0$ thus may be seen as a kind of compression gap, which quantifies how much compression is lost by neglecting temporal dependencies between the symbols~\cite{Schaub2012compression_gap}.

As an illustration, think of the movement in state space of a first order Markov process as a random walk on a graph.
If the graph is complete (including self-loops), $h_{KS} = H(Y_t)$, whereas if the graph is structured such that there will be preferred trajectories (e.g.\ a cycle) the gap between $h_{KS}$ and $H(Y_t)$ can be very large.
We may thus intuitively think of $I_0$ as measuring how structured the state space of the Markov process is.
Recall that for a Bernoulli process the state space may indeed be seen as a complete graph with self-loops, which means that $I_0$ is zero.
        
    \item We introduce the higher order Markov memory $\mathcal Q$:
    \begin{equation*}
      Q= I(Y_{t}, Y_{t+1}, \ldots \isep Y_{t-2}, Y_{t-3}, \ldots | Y_{t-1}) = \sum_{i>0}I_i
    \end{equation*}
    as the information contained in the terms beyond $I_0$.
    The higher order Markov memory can serve as a quantification of how close the process is to being first order Markov: $\mathcal Q$ is zero if and only if the process is (first order) Markovian.
    In general, a $k$-th order Markov process will have only $k$ non-zero terms in the reduced entrogram.
\end{enumerate}

\section{Block-structures, aggregated networks and their dynamical implications}
\label{sec:results}

So far we described general random processes.
We now focus on random processes $\mathcal Y$ obtained from coarse graining a Markov chain $\mathcal X$ on a network, and show how the dynamical properties of $\mathcal Y$, as expressed through its entrogram, reflect the structure of the network underlying $\mathcal X$.

Let us consider the first order Markov dynamics $\mathcal X$ associated with a random walk on a strongly connected network $\mathcal G=(V, E)$
defined by a set of nodes $V$ and a set of links $E$.
In this case we may simply identify the state space of our Markov process with
the set of nodes $V$. 
Stated differently, the process $\mathcal X$ is of the form $\mathcal X = ( \ldots, X_t,X_{t+1},\ldots)$, where the states $X_t\in V$ simply correspond to nodes in the network, and the transition probabilities between these states are encoded by the weights of the links in the graph.
As the process is first order Markov, the entrogram of the process $\mathcal
X$ comprises only a non-zero entry $I_0 = I(X_t\isep X_{t-1}, X_{t-2}, \ldots)$. 

\subsection{Higher order memory induced by projections --- the entrogram of a coarse grained Markov process}

\begin{figure}[htpb]
  \centering
  \includegraphics[width=0.6\linewidth]{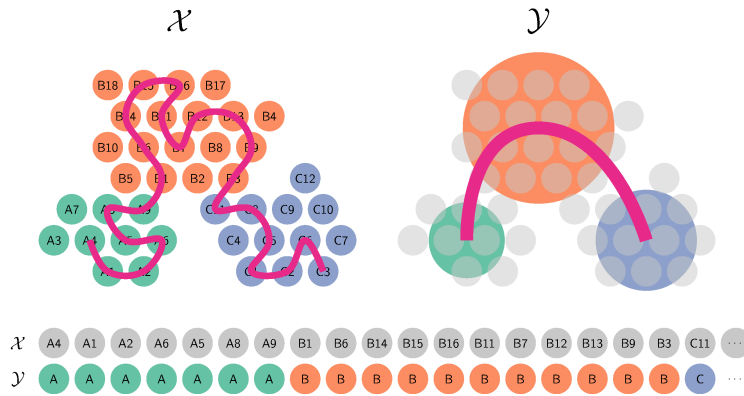}
  \caption{Projection of the Markov dynamics $\mathcal X$ to a partition
    $\mathbb P$.
    The Markovian process $\mathcal X$ is defined on the state space of nodes
    (upper left). The magenta path represents, e.g., the random walker path
    over the nodes on the network.
    It is projected step-by-step to the state space of blocks defined by the
    partition $\mathbb P$ (and characterized by colour codes on the upper right).
    On the bottom, a schematic description of both $\mathcal X$ and $\mathcal
    Y$ is given.
    Crucially, the new dynamical process $\mathcal Y$ is typically not a first order Markov process.
  }%
\label{fig:fig_clustering}
\end{figure}

Let us now consider a partition $\mathbb P$ of the nodes into non-overlapping blocks, and record the sequence of blocks traversed by the random walker (see Figure~\ref{fig:fig_clustering}). 
It is important to realize that the \emph{projected} dynamics $\mathcal Y = (\ldots,Y_t,Y_{t+1},\ldots)$, where $Y_t$ is the projection of $X_t$ onto the partition $\mathbb P$
(also called \emph{coarse grained}, \emph{aggregated}, or \emph{lumped}), is typically non-Markovian.
Further, as the projection is not a bijection, the total information contained in the entrogram of the projected dynamics is decreased, which follows simply from the information processing inequality~\cite{cover2012elements}.
Intuitively, as the projected paths necessarily cannot contain more information than the original, fine-grained paths, the total information contained in the projected process has to be smaller than the information contained in the original process.

Note that while the only non-zero entry of the entrogram of the original first
order process $\mathcal I^\mathcal{X} = [I_0^\mathcal{X}, \ldots ]$ will be
$I_0^\mathcal{X}=I(X_t \isep X_{t-1}, X_{t-2}, \ldots)$, the new entrogram $\mathcal{I}^\mathcal{Y}$ typically contains more than one non-zero component, and may even comprise infinitely many non-zero elements.
Since the states (symbols) of the new process $\mathcal Y$ are simply the blocks of the original graph, some quantities of the entrogram can be immediately interpreted in structural terms.
For example, if the new symbols of $\mathcal Y$ are uniformly distributed, $H(Y_t)$ is the logarithm of the number of blocks $K$.
Accordingly $H(Y_t)$ is less than $\log_2(K)$ if the steady state is unequally distributed.

As a side note, let us notice that the entropy rate needed to construct our entrograms, can be computationally estimated to arbitrary accuracy, by leveraging the doubly converging approximation~\cite{cover2012elements}:
\begin{equation}
H(Y_t|Y_{t-1}, \ldots, Y_{t-k+1}, X_{t-k}) \leq h_{\KS} \leq H(Y_t|Y_{t-1}, \ldots, Y_{t-k+1}, Y_{t-k}).
\end{equation}
This renders the entrogram effectively computable, at least for moderately sized networks. 
In the following we will thus focus on the conceptual ideas rather than the algorithmic aspects of computing these quantities efficiently for large networks. 
This will be a topic of future research.

\paragraph{Example 1. Preserving the Markov property: lumpable Markov chains.}

We may be interested in finding a projected dynamics that preserves the Markov property of the original, non-projected process.
As we will see in the next section, this implies that the dynamics captured in the lower-dimensional model correspond exactly to some subspace of the original dynamics.

Let us assume we can find a partition $\mathbb P$ of our state space, such that the process $\mathcal Y$ has the first order Markov property.
This coarse grained dynamics may thus simply be viewed as a random walk from block to block, described by a block-level transition matrix. 
The new system description thus has a smaller dimensionality ($\mathcal{Y}$ has less states than $\mathcal{X}$).
As the reduced system is still Markov, it has the desirable feature that it provides a simplified description of the dynamics which is exact for the specific coarse graining induced by $\mathbb P$ (see also Section~\ref{sec:community_or_lumbability_cases}).
In other words, within our projection the reduced model is in fact not an approximation, but captures the exact dynamics of the larger model.
However, other aspects of the dynamics are projected out, as the reduced model cannot capture all the information of the original model.
This provides us with an orthogonal decomposition of the dynamics into those captured by the reduced model and those exactly orthogonal to the model, which can be useful in practice.
Note that while we have required our projection to be Markov, we have not imposed any constraints on the entropy rate of the projected dynamics.
The entropy rate may be large or small irrespective of whether the process is Markov or not as we will see in next sections.

\paragraph{Example 2. Capturing time-scale separations: graphs with dense sub-clusters}
Let us now consider a network partitioned into groups of nodes (clusters), such that there are very few inter-group links and many intra-group links.
If there is a well defined block structure and the partition is chosen accordingly, the projected dynamics will have a very small entropy rate $h_\KS$.
In the coarse grained process, there will be few transitions between the blocks and the dynamics will mostly stay within the same block, leading to a chain of repeated symbols.

Note that the entropy rate of the projected dynamics is indicative of the probability, $p_\text{esc}$, to escape from a block.
For instance, if there are $K$ well-mixed (dense) clusters with a small probability $p_\text{esc}$ of escape at every time-step, then the entropy rate will be approximately 
\begin{equation*}
h_{KS} \approx h(p_\text{esc})+p_\text{esc} \log (K-1), 
\end{equation*}
where $h(x)=-x \log x - (1-x) \log (1-x)$ is the binary entropy function.
We may further gain some insight on the first entry of the entrogram $\mathcal E$ in this case, which will be approximately equal to $\log (K)$, if there are $K$ relatively equal-size clusters. 
As the entrogram entries are non-increasing, a small escape probability also implies small contributions from higher order terms, i.e., the projected process can be close to Markovian (though not exactly).
This latter fact follows alternatively from the results of Simon and Ando~\cite{simon1961aggregation}.

The above scenario may be viewed as analogous to some of the well-known time scale separation driven models in statistical physics, albeit here in a discrete setting.
For instance, microscopic physics is highly random over a huge state space. However, in many circumstances one can aggregate the state space to a few macroscopic variables such as total energy, average pressure, etc., that obey almost deterministic laws (small entropy rate), while the fast microscopic noise is projected out.

Interestingly, as we will discuss in the next section, there may exist many competing partitions which result in a projected process $\mathcal Y$ that is almost or even exactly Markovian, without a time-scale separation.

\subsection{Equitable partitions and lumpable dynamics}
Let us consider a weighted network with adjacency matrix $A$, out-degree vector $\mathbf{d}=A \mathbf{1}$ and corresponding diagonal matrix $D = \text{diag}(\mathbf{d})$.
For such a network, let us define the probability transition matrix of a random walk as $M=D^{-1}A$.
Note that each row sums to one and describes the outflow of probability from the corresponding node to other nodes in the network.
  
Any partition $\mathbb P$ of such a network can be characterised by an indicator matrix $P$, where $P_{jk} = 1$ if node $j \in V$ belongs to block $k$, and 0 otherwise.
We say a partition $\mathbb P$ is equitable if the out-flow of probability from every node to the different blocks depends only on the block containing the node, not the node itself.
Stated differently, an equitable partition satisfies: 
\begin{equation*}
MP= P M^\text{L},
\end{equation*}
where the block-to-block transition matrix $M^\text{L}$ is obtained by aggregating the transition matrix $M$ as
\begin{equation*}
M^\text{L} =  P^{+} M P,
\end{equation*}
where $P^+={(P^T P)}^{-1} P^T$ is the (Moore-Penrose) pseudo-inverse of $P$, which effectively acts as a cell-averaging operator~\cite{schaub2016synchro}.

The above equation implies that the set of vectors $\{v | v = Px, x\in \mathbb R^k\}$ that are constant on every group of the equitable partition spans an invariant subspace. 
It follows, that $M$ must have $K$ eigenvectors, which are constant on every group.
The significance of this fact is that if we aggregate the network according to the partition $\mathbb P$, then the resulting dynamics will remain exactly Markovian.
In other words, we obtain an exact reduced model for those eigenmodes that are constant on each cell, and an EEP thus precisely fulfills the criteria of our above discussed Example 1.
See Ref.~\cite{schaub2016synchro} for a related discussion for consensus dynamics.
It is important to remark here that the eigenvectors that are associated with an EEP do not have to correspond to the slowest modes in the system, i.e., do not have to be aligned with a time-scale separation in the system.
We will see examples of this in the following sections.

We pause here to highlight some parallels of the above discussion with some algorithmic approaches considered in the literature for (structural) graph partitioning.
Note that stochastic block models may be seen as a particular stochastic version of the notion of an equitable partition, where nodes within a same block are stochastically identical and share the same expected fraction of links towards each block. 
Indeed for a network generated by a stochastic blockmodel with $K$ blocks that is not too sparse (see also Ref.~\cite{Abbe2017} for a review on stochastic blockmodels and technical details), there will be a set of $K$ eigenvectors of a suitably chosen matrix, such as the adjacency matrix or the Laplacian, that will be \textit{approximately} constant on each of the $K$ groups.
Indeed most spectral methods used for finding the latent groups in stochastic blockmodels are centered around this fact~\cite{rohe2011spectral,Lei2015}.

In contrast, classical Laplacian based spectral clustering, arising from a relaxation of a (normalized) cut minimization problem, may be interpreted as identifying the slow modes in a diffusive dynamics~\cite{vonLuxburg2007}. The same is true for spectral coarse graining~\cite{GfellerLosRios2007spectralPRL}.
Hence, clusters found in this way tend to be well aligned with a (possible) time-scale separation in the system, but the reduced graph will typically not correspond to a perfectly equitable partition; and an associated projected diffusion dynamics would thus not be Markovian.

There can thus be different dynamically interesting partitions present in the system: some partitions may yield an almost perfectly Markovian description, but are not aligned with a separation of time-scales into slow and fast modes; other partitions may capture this time-scale separation better, but may lead to a non-Markovian description with a high order memory in the projected dynamics.
This tension between obtaining a projected dynamics that is Markov on one side, and a partition that captures a separation of time-scales in the dynamics on the other side, is illustrated by the following set of examples.

\section{Capturing different dynamical modes: core-periphery vs.\ cluster structure}%
\label{sec:community_or_lumbability_cases}

As discussed above different partitions may capture different aspects of the dynamics.
This issue can be well illustrated when examining networks in which both a core-periphery structure as well as a partition into dense clusters might be of interest.

\subsubsection*{An illustrative toy example: the bow-tie graph}%
\label{sssec:bowtie}

\begin{figure}[bt!]
  \begin{center}
      \includegraphics{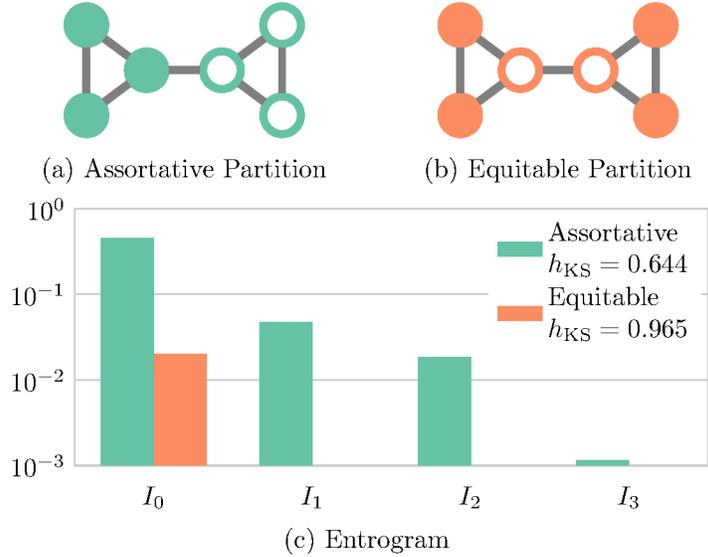}
  \end{center}
\caption{The entrogram of assortative and equitable partitions of the bow-tie
  graph (c).
  In the above graph, the left partition (a) represents a $2$-block partition
  akin to assortative cluster structure (where nodes from different blocks are marked with empty and filled symbols).
  This partition is approximately aligned with the dominant eigenvector.
  On the other hand, the partition in (b) is equitable, and is thus  spanned exactly by a set of (non-dominant) eigenvectors of the graph.
  In this case the projected dynamics is known to be Markovian.
}%
\label{fig:bowtie}%
\end{figure}

To gain some intuition let us initially consider the example of the bow-tie graph as illustrated in Figure~\ref{fig:bowtie}.
This graph may be partitioned in two different ways, which are essentially orthogonal to each other.
First, there is the assortative partition shown in Figure~\ref{fig:bowtie}a, which splits the graph into two triangles connected by a single link.
Second, there is the partition that splits the graph into the two
 central core nodes, acting as a bridge for the dynamics, and the four peripheral nodes (Figure~\ref{fig:bowtie}b).
Interestingly, this second partition is an equitable partition.

Both of these partitions capture quite different modes of the dynamics.
The assortative partition, captures the (relatively) slow transitioning of the diffusion between the two triangles.
In contrast the equitable partition captures the (relatively) fast transitioning between the core and the peripheral nodes.

In the assortative case, we observe a small entropy rate, indicating a predictable dynamics, which stems from the fact that in most cases there will be no transitioning between the blocks.
However, the total excess entropy of the projected dynamics is relatively large with substantial higher order memory terms present, due to the fact that the internal structure of the triangles has been projected out.

In contrast, the projected process derived from the equitable partition behaves more randomly, i.e., has a larger entropy rate $h_\KS\sim0.965$, indicating that there is an unpredictable alternance between the core and the peripheral nodes as we would expect.
Due to this large entropy rate, the information $I_0=H(Y_{t}) - h_{\KS}$ is relatively small.
Yet, as can be seen in Figure~\ref{fig:bowtie}c, the core-periphery split
corresponds to an equitable partition and thus the  entrogram
corresponds to a single bar $I_0$ and zero higher order Markov memory $\mathcal Q$, indicating a perfectly Markovian dynamics.

\subsubsection*{Core-periphery vs.\ assortative structures}
\begin{figure}[tb!]
\begin{center}
    \includegraphics{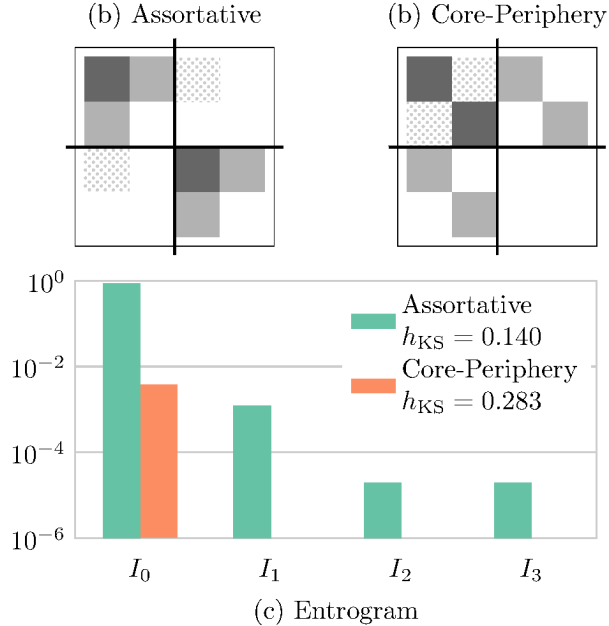}
\end{center}
\caption{Competing Partitions.  The same adjacency matrix is partitioned in
  (a) an assortative structure and (b) a core-periphery structure.
  The  entrograms of both partitions are compared in (c).
  While the assortative structure has a higher first bar, the values
  corresponding to higher order Markov effects remain non negligible.
  On the other hand the core-periphery structure exhibits a perfect first order
  Markovian dynamics.
  For the results shown in (c) a network of size $N=1000$ and rewiring
  ratio $f=0.011$ was used.
}%
\label{fig:ccp}%
\end{figure}

We can extend the bow-tie example by considering a graph with an adjacency matrix as depicted in Figure~\ref{fig:ccp}.
Here, the darker blocks represent higher density of links while white areas
contain no links.
The network can be assembled from two sub-networks with core-periphery structure.
Within each such subcomponent each node has a constant degree and a constant proportion of links towards the core and the peripheral nodes.
Specifically, in this example the core and periphery nodes have degree 19 and 1, respectively.
The two components are then connected by rewiring
a fraction $f$ of the links within the two cores (dotted area).

Let us assume again that we want to partition this network into two blocks.
By construction, two possible configurations are of dynamical interest:
the first depicted in Figure~\ref{fig:ccp}a, corresponds to a partition with an assortative, cluster-like structure, aligned with the slow time-scales in the system. 
However, it also results in a projected dynamics with large excess entropy, and additional memory.
The second partition in Figure~\ref{fig:ccp}b amounts to a (sub-divided)
core-periphery structure and the projected dynamics has zero higher order Markov memory, but also higher entropy rate when compared to the assortative case.
While most definitions of community or cluster would favour the first partition, the second split into a core-periphery structure can be of interest in a number of scenarios, as outlined in the work by Rombach et el.~\cite{rombach2014core, cucuringu_rombach_lee_porter_2016} and references therein.
From a dynamical perspective, one interesting aspect of the core-periphery
structure is the fact that the projected dynamics will be first order Markov, as the mesoscopic structure reproduces exactly the microscopic kinetics.
This can be of interest, for example, when a long simulation needs to be performed in a smaller, computationally tractable subspace. 
In this case a partition with non-zero higher order Markov memory will cumulate errors throughout the simulation as the original projected process $\mathcal Y$ will diverge from the Markovian simulation run on the projected topology. 
In such a scenario a partition leading to a coarse grained process with little or no higher order Markov memory $\mathcal Q$ would be preferable.

\subsubsection*{Graphs with Kronecker structure}
To generalize our above examples, let us consider a generic graph $\mathcal G$ with adjacency matrix $A$ and
$N$ nodes. 
We call the graph $\mathcal G$ the base graph.

We now construct the graph $\mathcal{\widetilde G}$ with the following Kronecker structure.
\begin{equation}
    \widetilde A = A \otimes (\text{Id}_K + \epsilon\mathbf{11}^T).
  \label{eq:kronecker}
\end{equation}
Here $\otimes$ denotes the Kronecker product and $\text{Id}_K + \epsilon\mathbf{11}^T$ is the $K$-by-$K$ identity matrix to which we add a non-negative perturbation of order $\epsilon$. 
We remark that this generic construction can be relaxed in many ways. 
For instance, we may use any adjacency matrix instead of the identity matrix, such as  $\left(\begin{smallmatrix} 0 & 1 \\ 1 & 0 \end{smallmatrix}\right)$, thereby planting a disassortative structure, or anti-communities. One may also consider a random perturbation $\epsilon$ rather than deterministic version.
Indeed, such type of graph models have appeared under the name Kronecker product graphs in the literature~\cite{Weichsel1962} and found applications as tractable model for graph generation and evolution~\cite{Leskovec2005,Leskovec2010}.

\begin{figure}[htpb]
  \centering
  \includegraphics[width=0.6\linewidth]{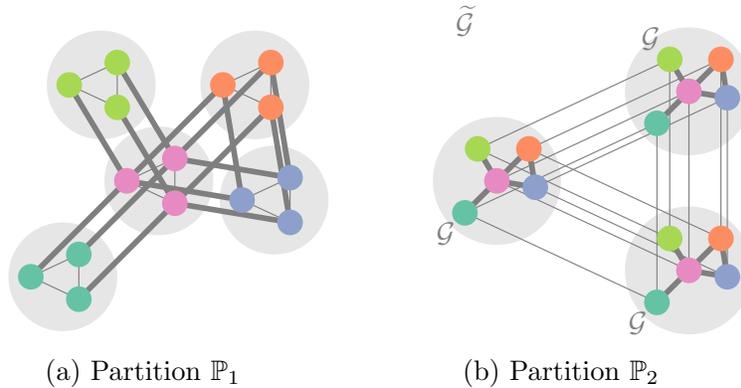}
  \caption{Graph with Kronecker structure.
    The Kronecker structure in~(\ref{eq:kronecker}) can be divided in two
    orthogonal partitions:
    the partition $\mathbb P_1$  (a) were each copy of the same node of $G$
    are arranged in the same block;
    and
    the assortative partition $\mathbb P_2$ (b) where
    each copy of the initial graph $G$ represents a densely connected
    cluster.
    Both partitions, despite providing different views of the same dynamics, are perfectly Markovian descriptions.
  }%
\label{fig:fig_kron}
\end{figure}

By construction, the adjacency matrix $\widetilde A$ admits two natural equitable partitions $\mathbb P_1$ and $\mathbb P_2$. 
In the first case each node in the base graph $\mathcal G$  will give rise to an
equitable block of nodes in $\mathcal{\widetilde G}$ (see $\mathbb P_1$ in Figure~\ref{fig:fig_kron}a).
The second equitable partition comprises the $K$ copies of the graph $G$ as individual blocks, which will be particularly relevant when $\epsilon$ is small (see $\mathbb P_2$ in Figure~\ref{fig:fig_kron}b).

The presence of these partitions is reflected in the spectrum of
$\mathcal{\widetilde G}$ as some simple analysis shows.
First note that the transition matrix of the graph $\mathcal{\widetilde G}$,
defined via $\widetilde M = {\widetilde D}^{-1}\widetilde A$,
where $\widetilde D=\text{diag}(\widetilde A\mathbf{1})$ has also a Kronecker product structure.
Namely, $\widetilde M$ is a product of the transition matrices $M$ for a random walk on $\mathcal G$, and the transition matrix $M_\epsilon$ corresponding to the transition matrix on the graph with adjacency matrix $\text{Id}_K + \epsilon\mathbf{1}\mathbf{1}^T$:
\begin{equation*}
  \widetilde M = M \otimes
  M_\epsilon
  \qquad
  \text{with}\ 
  M_\epsilon = {(1+\epsilon N)}^{-1}(\text{Id}_K + \epsilon\mathbf{11}^T)
\end{equation*}

The spectrum of the global random walk is therefore the Kronecker product of the individual spectra.
Thus the eigenvalues of $\widetilde M$ are given by all possible products $\lambda_M \lambda_{\epsilon}$, where $\lambda_M$ are the eigenvalues of $M$, and $\lambda_{\epsilon}$ are the eigenvalues of  $M_\epsilon$.
Note that $\lambda_{\epsilon}={(1+\epsilon N)}^{-1} = 1- \mathcal{O}(\epsilon)$ except for the dominant eigenvalue $\lambda_{\epsilon}=1$.
Further, the dominant eigenvalue of the transition matrix $M$ of the base graph $G$ is $\lambda_G=1$, and corresponds to the eigenvector $\mathbf{1}$ of all ones.
Therefore, for small enough $\epsilon$, the dominant eigenvectors of $\widetilde M$ are of the form $\mathbf{1} \otimes v_{\epsilon}$, for every eigenvector $v_{\epsilon}$ of $M_\epsilon$.
Hence, by construction these eigenvectors of $\widetilde M$ will be piecewise constant on each Kronecker copy of the base graph $\mathcal G$. 
Consequently, if we were to partition the graph according to the dominant eigenvectors, we will extract the partition $\mathbb P_2$ and not the split into the $N$-clique blocks, even though the latter is an alternative partition that leads to an exact lumping of the dynamics.
This case study can be extended to other similar topologies such as multiplex networks where the same node has a copy on several layers~\cite{moreno2013multiplex}.

\subsection{Core periphery vs.\ cluster structure --- real world examples}

Inspired by the previous  examples we test our tool on real networks where a similar dichotomy between assortative and core-periphery structure can be expected.
In order to obtain meaningful partitions for such networks we resort to the
embedded metadata to define the \emph{assortative} structure (which is also
aligned with what common community detection methods tend to find).
For the core-periphery structure we consider two alternative partitions: one defined by the set of nodes with highest degree, and the other by the set of nodes with the highest betweenness centrality.
While the former accounts for a structural separation present in most highly heterogeneous systems~\cite{rombach2014core}, the latter would separate the set of nodes which act as a bridge for the dynamics, promoting the diffusion of the dynamics between otherwise distant parts of the network.

We verify on two datasets how far these partitions are aligned with an information diffusion dynamics.
In order to assess the alignment of the partitions with some significant dynamical feature, we compare the entrograms to an ensemble of equivalent but randomly clustered systems, where nodes are assigned with equal probability to each cluster.
Figure~\ref{fig:realworld} displays the expected values (medians) of each $I_i$ computed on the system when projected to an ensemble of random partitions (grey bars in the figure).

\subsubsection*{The Karate Club}
A dichotomy between a cluster-like structure and a core-periphery arrangement is found in the Karate Club network of Zachary~\cite{zachary1977karateclub}.
Indeed, these two partitions are closely aligned with the optimal partition in the stochastic blockmodel and the degree corrected stochastic blockmodel~\cite{karrer2011dcsbm,Peel2017}, and appear also when trying to compress the network structure from a minimum description length perspective~\cite{Rosvall2007}.
The presumed cluster structure of the network, corresponding to the split of the Karate club into two factions, is moreover often used as a real world benchmark for community detection.

\begin{figure}[!ht]
\begin{center}
  \includegraphics[width=0.8\linewidth]{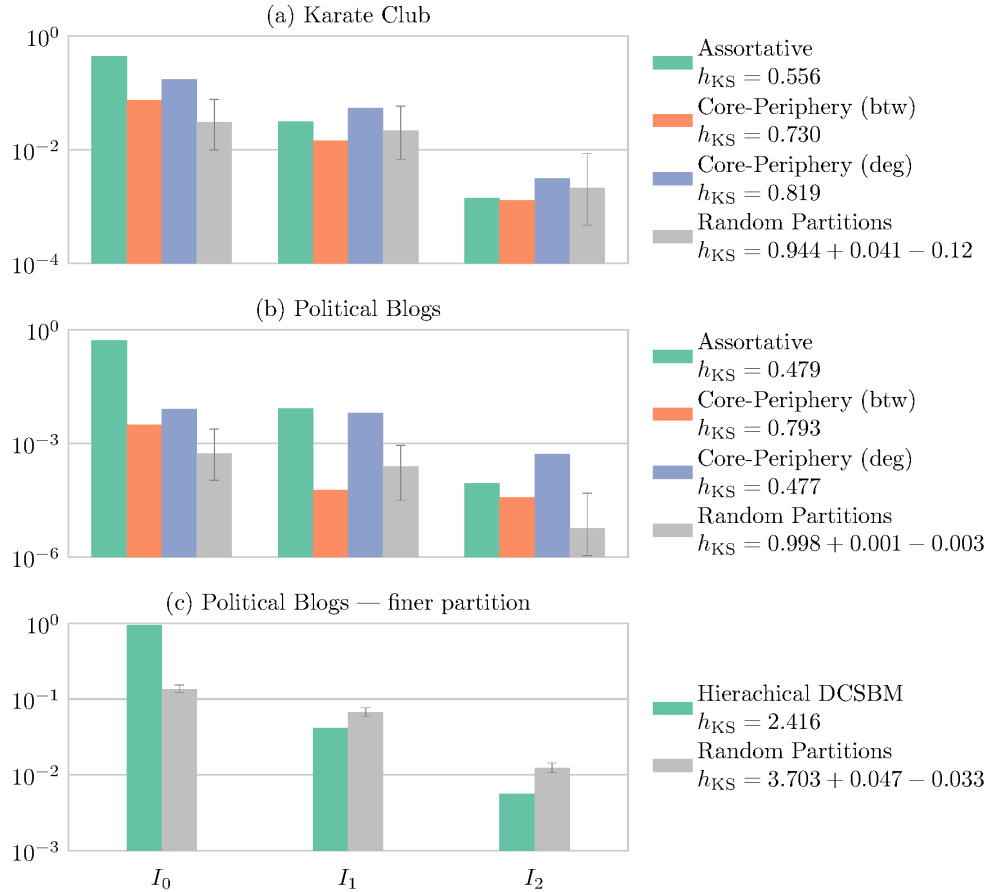}
\end{center}
\caption{Real world networks.
  The structure of the Karate club social network as examined by Zachary (a)
  and the political blog network described in~\cite{adamic2005politicalblogs} (b).
  Both networks can be divided according to two different block structures:
  a cluster-like, assortative partition, closely aligned with the eventual
  split of the Karate club or the political beliefs of the blog authors (green);
  and a core-periphery partition.
  In the latter the core can be defined as the set of nodes with higher
  degree (orange) or betweenness centrality (purple).
  The median of each index $I_i$, over an ensemble of randomized partitions,
  is reported in grey with its 5--95\% range
  demonstrating that both assortative and core-periphery
  structures are not mere fluctuations in the partition ensemble. 
  A finer partition in 15 clusters of the political blog network (c)
  exhibits smaller than expected higher order Markov memory, suggesting no
further meaningful substructure.
}%
\label{fig:realworld}%
\end{figure}

In Figure~\ref{fig:realworld}a we can see that in the both datasets the assortative and the degree-based core-periphery partition  are indeed capturing relevant features of the dynamic with higher than average $I_0$, with the former exhibiting the most predictable projected dynamics. 
The entropy rate values support this classification.
In contrast, the entrogram of the betweenness-based core-periphery partition  is indistinguishable from the entrogram of a random partition.

\subsubsection*{Political Blogs}
We now look at the social network of political blogs characterized in~\cite{adamic2005politicalblogs}.
The nodes of this network correspond to web blogs on US politics. 
The blogs are connected to each other in the network, if there is a hyper-link pointing from one blog to the other. 
In addition, each blog has been classified as being aligned with the Democratic or Republican party.

As in the previous examples, the system can be divided into an assortative
partition (characterized here from metadata~\cite{adamic2005politicalblogs})
or a core-periphery structure, in which the 240 (270) nodes with higher
degree (betweenness) centrality form a separate group.
The entrogram in Figure~\ref{fig:realworld}b exhibits a slightly different
pattern compared to the Karate Club case.

On the one hand, similar to before, the assortative structure has higher predictability as confirmed by the smaller entropy rate $h_\KS$ and higher $I_0$ compared to both the
core-periphery partitions.

On the other hand, in contrast to the previous case the higher order Markov memory $Q$ exhibits very different patterns.
While the betweenness-based core-periphery partition has a small higher order Markov memory $Q^\text{C-P} = 9.6 \cdot 10^{-5}$, capturing the Markovianity of the diffusion dynamics as in the example of Figure~\ref{fig:ccp}, the assortative partition and the degree-based core-periphery partition display large values for the higher order Markov memory $Q^\text{Ass} = 0.0083$ and $Q^\text{C-P} = 0.0067$, confirmed by the comparison to the expected values for a random bipartition of the system.
In particular the degree-based core-periphery structure exhibits a remarkably slow decay of memory, unlike the betweenness-based structure.

The larger than expected higher order Markov memory $Q$ exhibited by the assortative and degree based core-periphery clustering provides an indication that
different parts of a cluster may have different behaviours, in terms of
the sequence of clusters they will visit. For instance, if there are two dynamically
signiﬁcant partitions that overlap, then only a reﬁnement compatible with both of these partitions is likely
to be Markovian. The existence of a reﬁnement of the two-group partition of the political blog network has
also been suggested by Peixoto in~\cite{peixoto2014hierarchical}.
To investigate this further we consider a hierarchical partition.
The entrogram in Figure~\ref{fig:realworld}c depicts the results for a finer partition as found by the hierarchical stochastic block model algorithm in~\cite{peixoto2014hierarchical}.
The dynamical system divided into 15 clusters exhibits a smaller than expected higher order Markov memory $Q$, suggesting that no further dynamically meaningful substructure is to be found.

\section{Discussion}% 
\label{sec:ccl}

In this paper we introduce the entrogram, a tool inspired from the Computational Mechanics, as a way to summarize dynamical (Markov) processes, whose state space can be naturally identified by a network.
In the context of network analysis the entrogram compactly describes the memory properties of a random walk projected onto a partition.
The entrogram thereby provides a useful tool to analyse the merits of different putative partitions.
As the entrogram highlights, there are multiple dynamical features according
to which one could try to find a \emph{good} partition.
For instance, we may try to maximize the excess entropy (avoiding Bernoulli dynamics), or keeping the dynamics as close to a first order Markov process as possible, i.e., avoiding higher order memory in the coarse graining.

The examples discussed in this paper show how a dynamical perspective can lead to several different quality criteria for choosing a partition for aggregating a Markov process.
Depending on the partition, different aspects of the original dynamics will be preserved.
However, each partition offers only a partial, compressed view of the dynamics and thus there won't be a unique preferable partition in general that offers a favourable compression of the state space while preserving all aspects of the dynamics.

We believe that the entrogram can be a useful tool also in other scenarios, where an implicit or explicit dynamical description of a system is given: for instance, in analysing time-traces of Markov, higher-order Markov, or non-Markovian processes on networks~\cite{rosvall2014memory,delvenne2015diffusion}, or even in non-stationary processes~\cite{horvath2014spreading}.
As a natural next step one may look for efficient algorithms to search for the partitions inducing good or optimal coarse graining properties, or for a model-selection technique that delivers an appropriate number of blocks.   

\section{Acknowledgements}

The authors thank Leto Peel for fruitful discussions.
This work was supported by ARC (Federation Wallonia-Bruxelles) on Big Data Models and Methods,
IAP (Interuniversity Attraction Pole, Belgian Science Policy Office) DYSCO on Dynamical Systems, control and Optimization, a MOVE-IN Fellowship (M. Faccin), and the European Union's Horizon 2020 research and innovation programme under the Marie Sklodowska-Curie grant agreement No 702410 (M. Schaub); the results presented here reflect only the authors' views and the funders are not responsible for any use that may be made of the information contained in this article.

\bibliographystyle{unsrt}
\bibliography{biblio}

\end{document}